\documentstyle[12pt,epsfig]{article}

\begin{document}

\begin{center}
 {\Large \bf Vector-meson magnetic dipole moment effects in
radiative $\tau$ decays}\\
\vspace{1.3cm}

{\large G. L\'opez Castro and G. Toledo S\'anchez}

 {\it Departamento de F\'{\i}sica, Centro de Investigaci\'on y de
Estudios}\\ {\it Avanzados del IPN, Apdo. Postal 14-740, 07000 M\'exico,
D.F., M\'exico}

\end{center}
\vspace{1.3cm}
\begin{abstract}
 We study the possibility that the magnetic dipole moment of
light charged vector mesons could be measured from their effects in
$\tau^- \rightarrow V^-\nu_{\tau}\gamma$
decays. We conclude that the energy spectrum and angular
distribution of photons emitted at small angles with respect to vector
mesons is sensitive the effects of the magnetic dipole moment. 
Model-dependent contributions and photon radiation off
other electromagnetic multipoles are small in this region. We also
compute the effects of the magnetic dipole moment on the integrated rates
and photon energy spectrum of these $\tau$ lepton  decays.
 \end{abstract}

 \vspace{1.3cm}
PACS number(s): 13.35.Dx, 13.40.Em, 14.40.Cs

\newpage

\begin{center}
\large \bf 1. Introduction
\end{center}

   The magnetic dipole moment ($\mu$) and electric quadrupole moment
(${\cal Q}$) of vector mesons ($J^P=1^-$), and more generally of spin-one
elementary and hadronic particles, have not been measured yet. In the case
of vector mesons, even upper bounds have not been reported by experiments
up to now. The very short lifetime of vector mesons do not allow to use
vector-meson--electron scattering (as in the case of the deuteron) or the
spin-precession technique \cite{bmt} (as useful for hyperons) to measure
their electromagnetic static properties. Instead, only photoproduction
experiments or radiative decays may eventually be able to measure these
multipoles \cite{kpz}. In this paper we study the sensitivity of $\tau^-
\rightarrow V^- \nu_{\tau} \gamma$ decays to the magnetic dipole moment of
the vector meson $V^-$ and identify an observable associated to this decay
that could provide a measurement of this important property.

Measurements of the magnetic dipole and electric quadrupole moments of the
$W$ gauge bosons would provide a significant test for the gauge
structure of the standard model of particle interactions, while the
corresponding multipoles of vector mesons contain information on the
structure of these hadrons and, ultimately, about the dynamics of strong
interactions.
At present, more restrictive bounds on
the magnetic dipole and electric quadrupole moments of the $W^{\pm}$
gauge bosons are being provided
from experiments at the LEPII \cite{lepII} and Tevatron colliders
\cite{tev}. These bounds are
consistent with the predictions based on the standard model, while 
the corresponding experimental information for vector mesons is absent.
The simplest of the spin-one systems whose multipoles have been
measured with good precision is the deuteron ($J^P=1^+$): $\mu_{D}=
0.85741\mu_N ,\ {\cal Q}_{D} = 0.2859\ {\rm fm}^2$ \cite{deut}.
This information confirms the picture that the deuteron
can be viewed as a weakly bound state of a neutron and a proton whose
electromagnetic properties can be understood, for instance,  in terms of a
model for low energy interactions of baryons and mesons \cite{wise}.

   In a previous paper we have considered the possibility to measure the
magnetic dipole moment of light vector mesons $\rho^{\pm}$ and $K^{*\pm}$
by looking at the energy and angular distribution of photons emitted in
their two-body radiative decays ($V^{\pm} \rightarrow P^{\pm} 
P^0 \gamma$, where $P$ denotes a pseudoscalar meson)\cite{nos}.
We have found that this
observable is more sensitive to the effects of the magnetic dipole moments
than their corresponding integrated rates \cite{dbl}. Indeed, measurements
of the photon spectrum in the  kinematical
region where this observable is dominated by the emission off the
magnetic dipole moment, would provide a determination of this property
within $\Delta \mu = \pm 0.5$ (in units of $e/2m_V$) if the photon 
spectrum is measured with a precision of around 25 \% \cite{nos}.

  In the present paper we are concerned with the possibility of measuring
the magnetic dipole moment of the vector mesons $\rho^{\pm},\ K^{*\pm}$ in
the radiative decays of the $\tau$ lepton, namely $\tau^- \rightarrow V^-
\nu_{\tau} \gamma$ (henceforth our discussion will be focused on the
$\rho^-$ vector meson, but everything applies to the $K^{*-}$ meson with
the appropriate changes in flavor indices). Among the motivations for this 
study we
find the following: $(a)$ charged vector mesons are produced in a
clean way through  the decay  $\tau^- \rightarrow V^- \nu_{\tau}$
and the $\rho^-$ accounts for almost 25 \% of all $\tau$ decays
\cite{pdg}, therefore the corresponding decay with an accompanying photon
can be expected at a fraction of a percent level and, $(b)$ there are
several recent studies of the electromagnetic vertex of light vector
mesons that provide a computation of their electromagnetic multipoles
\cite{hecht, hawes}. 

  An implicit shortcoming of dealing with vector mesons (and unstable
particles in general) is
that the separation of the production and decay processes of a resonance
are in general model-dependent. Actually, only the full S-matrix amplitude
that involves de production and decay of an intermediate resonance is
physically meaningful. Therefore, the consideration of truncated   
processes 
such as the radiative decays $V^+ \rightarrow P^+P^0 \gamma$
\cite{nos,
dbl} or the radiative production of a resonance as in the present study
necessarily involves a factorization approximation. The effects of the
vector meson magnetic dipole moment in the full production and decay
processes (for example,  $\tau^- \rightarrow \rho^- \nu_{\tau} \gamma
\rightarrow \pi^- \pi^0 \nu_{\tau} \gamma$ in the present case) and the
difficulties associated to a gauge-invariant formulation of the
electromagnetic vertex of an unstable vector particle \cite{gauge} goes
beyond the purpose of the present paper \cite{nos2}. 

  This paper is organized as follows. In section 2 we analyze the effects
of the magnetic dipole moment of the vector mesons $\rho^-$ or $K^{*-}$ in
the energy and angular spectrum of photons in $\tau^- \rightarrow
(\rho^-,\ K^{*-}) \nu_{\tau} \gamma$ decays. These effects in the
corresponding integrated rates and photon energy spectra are given in
section 3. Conclusions are
summarized in section 4 and the explicit formulae for the differential
decay rates of $\tau$ lepton radiative decays are provided in the
Appendix.

\

\begin{center}
\large \bf 2. Effects in the energy and angular distribution of photons
\end{center}

   Let us start by setting our notations. The relevant electromagnetic
vertex of the charged vector meson $V$ is defined as follows (the flow
of momenta and Lorentz indices are chosen according to $V^+_{\alpha}(
q_1)
\rightarrow V^+_{\beta}(q_2) \gamma_{\mu}(k)$):
 \begin{equation}
\Gamma^{\alpha \beta \mu} = g^{\alpha
\beta}(q_1+q_2)^{\mu}\alpha(k^2) +
(k^{\alpha}g^{\mu\beta}-k^{\beta}g^{\mu\alpha}) \beta(k^2) +
(q_1+q_2)^{\mu} k^{\beta} k^{\alpha}\gamma(k^2) \ .      
 \end{equation}
The form factors $\alpha(k^2),\ \beta(k^2)$ and $\gamma(k^2)$ at
$k^2=0$ are related (see for example \cite{bm}) to the electric charge $q$
(in units of $e$),  
the magnetic dipole moment $\mu$ (in units of $e/2m_V$) and electric
quadrupole moment ${\cal Q}$ (in units of $e/m_V^2$) through the
relations $\alpha(0)=q=1$, $\beta(0)= \mu$ and $\gamma(0)= (1-\mu-{\cal
Q})/2m_V^2$.

The value $\beta(0)=2$ for the giromagnetic ratio can be considered as the
canonical value \cite{bm} and it appears in a natural way in a model where
the isomultiplets of vector mesons couple through non-Abelian Yang-Mills
terms, in an analogous way to the $W^{\pm}$ gauge boson
couplings in the stantard model . Also, according to several authors the
canonical value $\beta(0)=2$ can be considered as the gauge condition for
spin-one particles \cite{g2}. However, the Yang-Mills nature of the
vector mesons is far from being proved and therefore substancial
deviations from the canonical value may be expected. For instance, the
estimates of Refs. \cite{hecht, hawes}, indicate that the magnetic moments
of the light  vector mesons are in the range $\beta(0) = 2.2 \sim 3.0 (
2.37)$ for the $\rho^- \ (K^{*-})$ mesons.

   Following the discussion in our previous paper \cite{nos}, we do not
consider the effect of the form factor $\gamma(k^2)$ because it enters the
amplitude at the same order in the photon energy as does the  
model-dependent contributions \cite{low} (for example $\tau^- \rightarrow 
\pi^- \nu_{\tau} \rightarrow \rho^-\gamma \nu_{\tau}$ in the present
case), and we expect them to be suppressed
for the special kinematical configuration to be discussed below (see
also Ref. \cite{nos}).

   The four-momenta and the polarization four-vectors ($\epsilon^*_{\mu}$
and $\eta^*_{\nu}$)
 of the vector particles are chosen as  $\tau^-(p) \rightarrow V^-(q, \
\eta) \nu_{\tau}(p') \gamma(k,\ \epsilon^*)$. Up to terms of order
$k^0$, the gauge-invariant amplitude for the process $\tau^- \rightarrow
\rho^- \nu_{\tau} \gamma$ is given by:
\begin{eqnarray}
{\cal M} &=& \frac{eg_{\rho}G_F V_{ud}}{\sqrt{2}} 
 \left \{ \overline{u}(p') O^\alpha u(p)
\left(\frac{q \cdot \epsilon^*}{q \cdot k}-\frac{p \cdot \epsilon^*}{p  
\cdot k}\right)\eta^*_\alpha \right. \nonumber\\
&& + \left.  \overline{u}(p') O^\alpha \frac{\not{k}
\not{\epsilon}^* }{2p \cdot k}u(p) \eta^*_\alpha
+ \frac{\beta(0)}{2q\cdot k}\overline{u}(p') O^{\alpha} u(p)  
\left( k_{\alpha}\epsilon^*_\lambda - k_\lambda
\epsilon^*_\alpha \right)\eta^*_\lambda \right.  \\
&&+ \left. \frac{m_\tau}{m^2}(1-\frac{\beta(0)}{2})\overline{u}(p')
(1+\gamma_5) u(p) \left(\epsilon^*_\lambda - \frac{q \cdot \epsilon^*}{q
\cdot
k}k_\lambda \right)\eta^*_\lambda  \right \} \nonumber
\end{eqnarray}
where $m_{\tau}$ and $m$ denote the masses of the $\tau$ lepton and the
vector meson, respectively. 
$G_F$ and $V_{ud}$ denote the Fermi constant and the
Cabibbo-Kobayashi-Maskawa mixing matrix element, respectively, while
$O^{\alpha}\equiv \gamma^{\alpha}
(1-\gamma_5)$. The constant $g_{\rho}$ denotes the $W-\rho^-$ strength
coupling. As stated by Low's theorem \cite{low}, this amplitude depends
only on the static electromagnetic properties of the particles, the
parameters of the non-radiative process and it is model-independent.

   It is interesting to look at the structure of the squared amplitude in 
terms of the independent Lorentz scalars of the process. If
we sum over polarizations of fermions and the vector meson we obtain (an
overall factor $(eg_{\rho}G_F |V_{ud}|/\sqrt{2})^2$ is omitted):
\begin{eqnarray}
\sum_{pols}|{\cal M}|^2&=&
\frac{4m_{\tau}^2}{r}\left |\frac{p\cdot \epsilon }{p \cdot
k}-\frac{q\cdot \epsilon}{q
\cdot k}\right|^2
\left\{(1-r)(1+2r) +4\frac{(p \cdot
k)^2}{m_{\tau}^4}\left(1-\frac{\beta(0)}{2}\right)^2
\right\}\nonumber\\
&&+8\epsilon \cdot \epsilon^* \frac{q \cdot
k}{rm^2}\left(1-\frac{\beta(0)}{2}\right)^2 (1 +r) \nonumber\\
&&+4\epsilon \cdot \epsilon^* 
\left\{
\left(-1+\frac{q \cdot k}{p \cdot k} \right)\left[1-\beta(0)\frac{p \cdot
k}{q \cdot k}
-\frac{1}{r}\left(1-\frac{\beta(0)}{2}\right) \right]^2 \right.
\nonumber\\
&&\left.+\left(1-\frac{1}{r}\frac{q \cdot k}{p \cdot
k}\right)\left[\frac{1}{r}\left(1-\frac{\beta(0)}{2}\right)^2 - \beta(0)
\right]\right. \\
&&\left. +\left(1-\frac{1}{r}\right)\left[
\left(\frac{\beta(0)}{2}\right)^2+\frac{q \cdot k}{p \cdot k}\right]
\right\}\ , \nonumber
\end{eqnarray}
where we have introduced the dimensionless quantity $r\equiv
(m/m_{\tau})^2$.
This squared amplitude satisfies the Burnett and Kroll theorem \cite{bk},
{\it i.e.} does not contain terms of $O(k^{-1})$. Eq. (3) takes a very
simple form when $\beta(0)=2$.

   When we sum over photon polarizations, the first term in Eq. (3)
becomes proportional to the infrarred factor (we will place in the rest
frame of the $\tau$ lepton):
\begin{equation}
\sum_{pols\ \epsilon} \left |\frac{p\cdot \epsilon }{p \cdot
k}-\frac{q\cdot \epsilon}{q \cdot  k}\right|^2 = \frac{|\vec{q}|^2
\sin^2\theta}{\omega^2(E-|\vec{q}|\cos \theta)^2}
 \end{equation}
where $\theta$ is the angle between the photon and the vector meson
three-momenta, $E$ ($\omega$) is the energy of the vector meson (the
photon) and $E=\sqrt{\vec{q}^2+m^2}$ in 
this frame. This term is suppressed for small
values of the angle $\theta$, therefore the photon radiation off the
electric charges of the $\tau$ lepton and the vector meson can be
suppressed for 
this angular configuration. In the same way, it can be shown that this
small angle configuration allows to suppress the interference terms of
$O(k^0)$ that comes from the interference between the quadrupole (or
model-dependent) terms with the emission off the charges of $\tau$ and
$V$. 

  Therefore, it becomes convenient to split the unpolarized
differential decay rate into two
terms. The first one, which is associated to the first term in Eq. (3),  
vanishes when $\theta=0,\ \pi$ (collinear photons). This term reduces to
the radiation emitted  
from the electric charges of the $\tau^-$ and $V^-$ when $\beta(0)=2$. The
second term (remaining terms in Eq. (3)), does not vanish for
collinear photons and contains a
sensitive dependence on the magnetic dipole moment of the vector meson. If
we denote these two terms with subscripts 1 and 2, respectively, we can
write the normalized (to the non-radiative decay rate
$\Gamma_{nr}=\Gamma(\tau^- \rightarrow V^- \nu_{\tau})$)
double differential decay rate as follows:
\begin{equation}
\frac{d\Gamma}{\Gamma_{nr}dxdy}=\frac{1}{\Gamma_{nr}}\sum_{\pm}
\left(\frac{d\Gamma^{\pm}_1}{dxdy}+\frac{d\Gamma^{\pm}_2}{dxdy}\right)
\end{equation}
where we have introduced the dimensionless variables $x=2\omega/m_{\tau}$
and $y=\cos\theta$. 
   The explicit expressions for the two contributions to the differential
decay rate are given in the Appendix. 

The normalized rates given above do
not depend on the $g_{\rho}$ coupling  and the CKM matrix
element. Their only dependence is on the magnetic dipole moment
$\beta(0)$. Following the previous discussion, in Figs. 1 ($\rho^-$ meson) 
and 2 ($K^{*-}$) we plot
the differential decay rates of Eq. (5) as a function of the photon
energy ($x$) for fixed values of the angle $\theta$ ($\theta=10^0\
(20^0)$ corresponds to the upper (lower) half of the plots) and three
different values of $\beta(0)$. The short-dashed line in the different
plots denotes the first term of Eq. (5), while the other lines refer to
the second term, for the different values of $\beta(0)$: $\beta(0)=1$ (
solid line), $\beta(0)=2$ (long-dashed line) and $\beta(0)=3$
(long--short-dashed line). 

     As is evident from these plots, the model-independent
terms of order $k^0$ dominate over the radiation due to the electric
charges (terms of $O(k^{-2})$ in the squared amplitude) in the region of
photon energies $0.2 < x < 0.5$ for the $\tau^-
\rightarrow \rho^- \nu_{\tau} \gamma$ decay and $0.2 < x < 0.4$ for the 
$\tau^- \rightarrow K^{*-} \nu_{\tau} \gamma$ case. Therefore, it is
precisely in this kinematical region where the 
measurement of the photon spectrum amy provide a
determination of the vector meson magnetic moment with a reasonable
accuracy.

The curves also
exhibit a dip at  $x=0.5667$ for the $\rho^-$  case and $x=0.4991$
for the case of $K^{*-}$ production. The position of this dip
is independent of $y$ and the value of $\beta(0)$. Furthermore, the dip
is more
pronounced when $\beta(0)=2$. At first sight this dip could be associated
to a {\it null} radiation amplitude \cite{msb}, however it can be checked
that  it corresponds to the special kinematical configuration where the
vector meson remains at rest. Indeed, the dominant contribution to the
differential decay rate is the one evaluated at the $E^-$ solution (see
the Appendix) and this rate vanishes when $v^- =0$ or, equivalently,
$x=1-\sqrt{r}$ ({\it i.e.}, the position of the dip in the photon energy
spectrum is independent of $y$ and $\beta(0)$)..

\

\begin{center}
\large \bf 3. Effects in the integrated rate and single photon spectrum.
\end{center}

  For completeness, we also provide the dependence of the decay rate
and the photon energy spectrum for
the process $\tau^- \rightarrow (\rho^-, K^{*-}) \nu_{\tau} \gamma$ upon 
the magnetic dipole moment of the vector meson. The general expression for
the branching ratio is
\begin{equation}
B(\tau \rightarrow V^- \nu_{\tau} \gamma; \omega > \omega_{min}) =
\frac{\alpha}{2\pi } \frac{r B(\tau^- \rightarrow V^-
\nu_{\tau})}{m_{\tau}^2(1-r)^2(1+2r) }
\int_{\omega_{min}}^{(m_{\tau}^2-m^2)/2m_{\tau}}d\omega 
\int_{E^{min}}^{E^{max}} dE
F(E,\omega)
 \end{equation}
where,
\begin{eqnarray}
F(E,\omega)&=& \frac{8\omega^2 (E^2-m^2) -2(A+2E\omega )^2}{\omega^2 A^2r}
C(\omega) \nonumber \\
&& + 8\left( 1+ \frac{A}{2m_{\tau}\omega} \right) \left\{ 1-\frac{1}{r} 
\left(1-\frac{\beta(0)}{2} \right) + \frac{2\beta(0) m_{\tau} \omega}{A}
\right \}^2 \nonumber \\
&& -8 \left( 1+\frac{A}{2rm_{\tau}\omega} \right) \left(
\frac{1}{r}\left(1-\frac{\beta(0)}{2} \right)^2 -\beta(0) \right) \\
&& +8\left(\frac{1}{r}-1\right) \left( \left(\frac{\beta(0)}{2} \right)^2
-\frac{A}{2m_{\tau}\omega} \right) + \frac{8A}{m^2}
\left(1+\frac{1}{r}\right) 
\left(1-\frac{\beta(0)}{2} \right)^2 \ , \nonumber
\end{eqnarray}
and
\begin{eqnarray}
A&=& m^2_{\tau}+ m^2-2m_{\tau}\omega -2m_{\tau} E\ , \nonumber \\
C(\omega)&=&  m_{\tau}^2(1-r)(1+2r) +
4\omega^2\left(1-\frac{\beta(0)}{2} \right)^2\ , \\
E^{max,min} &=& \frac{(m_{\tau}-\omega)(m_{\tau}^2-2m_{\tau}\omega+m^2)
\pm
\omega (m_{\tau}^2-2m_{\tau}\omega -m^2)}{2(m_{\tau}^2-2m_{\tau}\omega)}.
\nonumber 
\end{eqnarray}
and $\alpha\equiv e^2/4\pi \approx  1/137$ denotes the fine structure
constant.
Thus, the branching ratio for the radiative decays depends only upon the
minimun photon cutoff energy $\omega_{min}$ the experiment is able to
discriminate and also depends on the magnetic dipole moment of vector
meson $\beta(0)$.

  In Table 1 we display the branching ratios for $\tau^- \rightarrow
(\rho^-, K^{*-}) \nu_{\tau} \gamma$ decays for three different values of
$\beta(0)$ and four realistic values of the photon cutoff energy
$\omega_{min}$. The branching ratios of the non-radiative processes
$\tau^- \rightarrow V^- \nu_{\tau}$ were taken from Ref. \cite{pdg}. As
expected, the branching ratios in Table 1 are of order $10^{-2}$ with
respect to their non-radiative counterparts. We
observe that these branching ratios approach their minimum values when  
$\beta(0)$ is close to its canonical value $\beta(0)=2$. We observe
also that these ratios become more sensitive to the effect of $\beta(0)$
when $\omega_{min}$ increases. This happens because the radiation
due to the electric charges is more important at the lower end of
the photon energy spectrum, while higher electromagnetic
multipoles are more important for higher energies of the radiated photon. 

  Finally, we can also compute the effects of $\beta(0)$ in the photon
energy spectrum. The single photon energy spectrum is obtained from 
(6) after performing the integration only over $E$. The plots for
the normalized photon energy spectrum $(1/\Gamma_{nr}) d\Gamma/dx$, where
$x\equiv 2\omega/m_{\tau}$,  for values of $\omega \geq 100$ MeV are shown
in Fig. 3 (for $\tau \rightarrow \rho \nu_{\tau} \gamma$) and Fig. 4 (for
$\tau \rightarrow K^* \nu_{\tau} \gamma$). The solid, long-dashed and
short-dashed lines correspond, respectively, to $\beta(0)=1,\ 2$ and $3$.
These single photon spectra are more sensitive in the region of
intermediate photon energies. However, as in the case of the integrated
rates, this observable is not as sensitive to the effects of the
magnetic dipole moment as the double differential photon spectrum
discussed in section 2.

\begin{center}
\large \bf 4. Some conclusions.
\end{center}

   In this paper we have explored the effects of the magnetic dipole
moments of vector mesons on the $\tau^- \rightarrow (\rho^-, K^{*-})
\nu_{\tau} \gamma$ decays. Our main interest is on the energy and angular
distribution of the emitted photon. We observe that this differential
decay rate is sensitive to the effects of the magnetic dipole moment when 
the photon is emitted at small angles  with respect to the vector meson. 
This
effect is more important for intermediate photon energies where radiation
off the magnetic dipole moment dominates over the radiation due to the
electric charges and the electric quadrupole moment of vector mesons and
the other model-dependent contributions. For completeness, we have
computed also the effects of the magnetic dipole moment on the
corresponding branching ratios and photon energy spectrum of these decays
and found that they are less sensitive to these effects.

   Since vector mesons are detected through their decays into two
pseudoscalar mesons, the present study would require to reconstruct the
invariant masses of the two-pseudoscalar in a region close to the vector
meson mass. On the other hand, it requires to be able to detect
photons emitted at small angles with respect to the vector meson. Those
experimental capabilities would allow a measurement of the magnetic dipole
moments from the most favored region in the differential decay rate with a
reasonable accuracy.

\begin{center}
\large Acknowledgements
\end{center}

  We are grateful to A. Garc\'\i a and A. Queijeiro for useful 
discussions.

\

\begin{center}
{\bf Appendix}
\end{center}

  The expressions for the differential decay rates appearing in Eq. (5)
are:

\begin{equation}
\frac{d\Gamma^\pm_1}{\Gamma_{nr}dxdy}=\frac{2\alpha x(1-y^2)}
{\pi\left| (2-x)v^{\pm}+xy\right| (1-r)}
\frac{(E^{\pm})^3(v^{\pm})^4 }{(T^{\pm})^2} \nonumber 
\end{equation}

and 

\begin{eqnarray}
\frac{d \Gamma^\pm_2}{\Gamma_{nr}dxdy}&=&
\frac{2x\alpha E^{\pm}(v^{\pm})^2}
{\pi\left| (2-x)v^{\pm}+xy \right| (1-r)^2(1+2r)} \nonumber \\ 
& & \times \left[ \left(
(1-y^2)\left(\frac{xE^{\pm}v^{\pm}}{T^{\pm}}\right)^2
+\frac{1}{r}T^{\pm}(1+r)
\right)\left(1-\frac{\beta(0)}{2}\right)^2  \right. \nonumber \\
&&\left.+\frac{r}{x}\left(T^{\pm} +x \right)
\left[ 1+\frac{\beta(0)x} {T^{\pm}}
-\frac{1}{r}\left(1-\frac{\beta(0)}{2}\right)
\right]^2 \right. \nonumber \\   
&&\left.+\frac{1}{x}
\left(T^{\pm} + xr\right)\left[
\beta(0)-\frac{1}{r}\left(1-\frac{\beta(0)}{2}\right)^2\right] \right.
\nonumber \\
&& \left. +\frac{(1-r)}{x}\left(
x\left(\frac{\beta(0)}{2}\right)^2-T^{\pm})
\right) \right]\ .
\end{eqnarray}

In the above expressions, $ \alpha\equiv e^2/4\pi$ is the fine
structure constant, $\ v^{\pm}
=\sqrt{1-r/(E^{\pm})^2}$ and 
\begin{eqnarray}
T^{\pm}&=& 1+r-x-2E^{\pm}\nonumber \ , \\
E^{\pm} &=& \frac{1}{(x-2)^2-x^2y^2} \left\{ (2-x)(1+r-x)
\pm xy\sqrt{(1-r-x)^2-x^2r(1-y^2)}
\right\}\ .
\end{eqnarray}

\newpage
\begin{center}

\begin{tabular}{|l|c|c|c|c|c|}
\hline\hline
Channel &$\beta(0)$&$\omega_{min}$=15&30&60&90\\
\hline 
$\rho^-$&1&0.07618&0.05900&0.04268&0.03374\\
&2&0.07332  &  0.05614  &  0.03984 &  0.03094\\
&3&0.07981 &  0.06260  &  0.04616 &  0.03705\\
\hline 
$K^{*-}$&1&0.00246&0.00187&0.00131&0.00101\\
&2&   0.00243 &  0.00184  &  0.00128  & 0.00098\\
&3&   0.00267 &  0.00207  &  0.00151  & 0.00119\\
\hline 
\end{tabular}
\end{center}

\

Table 1: Branching ratios (in \%) for the $\tau^- \rightarrow (\rho^-,
K^{*-}) \nu_{\tau} \gamma$ decays as a function of $\beta(0)$ and the
lower cutoff photon energy $\omega_{min}$ (in MeV).

\newpage

\begin{figure}
\label{Figure 1}
\centerline{\epsfig{file=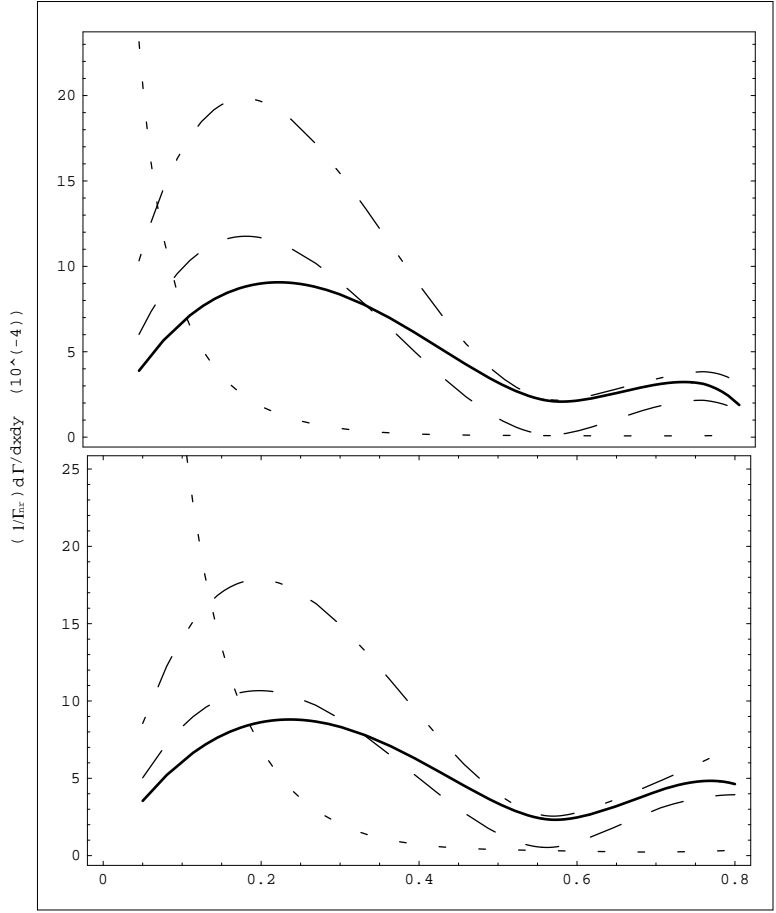,angle=0,width=4.0in}}
\vspace{-0.5in}
\caption{Differential photon spectrum for $\tau \rightarrow \rho \nu
\gamma$ decay. See description after Eq. (5).}

\end{figure}

\newpage

\begin{figure}
\label{Figure 2}
\centerline{\epsfig{file=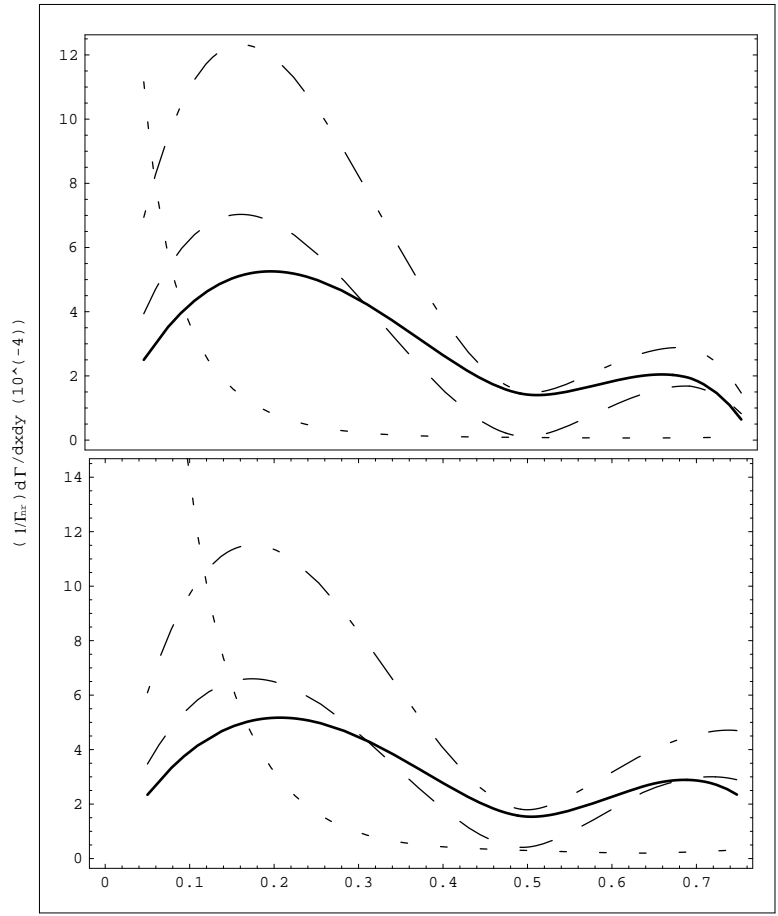,angle=0,width=4.0in}}
\vspace{-0.5in}
\caption{Same as in Fig. 1 for the $\tau \rightarrow K^* \nu \gamma$
decay.}

\end{figure}

\newpage

\begin{figure}
\label{Figure 3}
\centerline{\epsfig{file=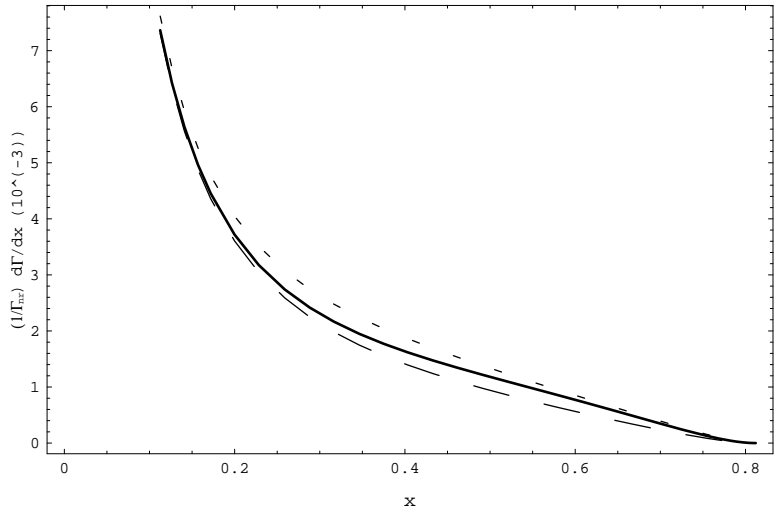,angle=0,width=4.0in}}
\vspace{-0.5in}
\caption{Normalized photon energy spectrum of the decay $\tau \rightarrow
\rho \nu \gamma$ for $\beta(0)=1$ (solid line), 2 (long-dashed) and 3
(short-dashed).}

\end{figure}

\begin{figure}
\label{Figure 4}
\centerline{\epsfig{file=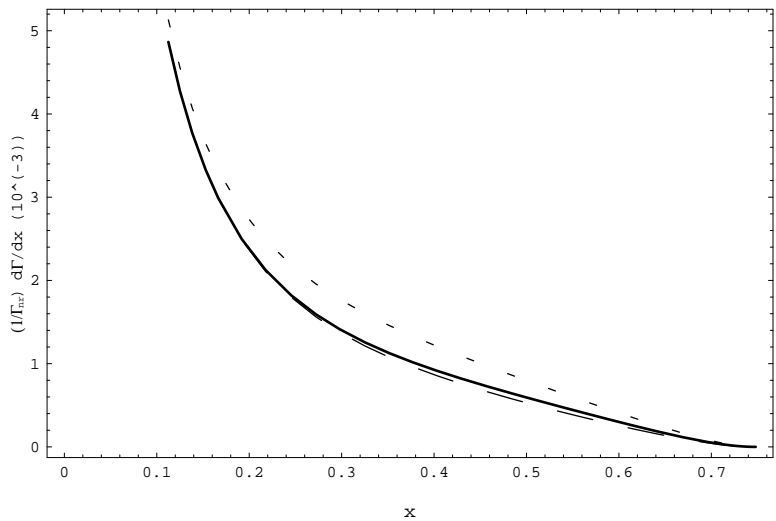,angle=0,width=4.0in}}
\vspace{-0.5in}
\caption{Same description as in Fig. 3 for the $\tau \rightarrow K^* \nu
\gamma$ decay.}

\end{figure}

\end{document}